\documentclass[useAMS,usenatbib]{mn2e}
\usepackage{epsf,palatino,url,graphicx,wrapfig,array,setspace,amssymb,fancyhdr,multirow,lscape,appendix,rotating,amsmath}

\def\apj{ApJ }
\def\aap{A\&A }
\def\aaps{A\&AS }

\def\pasp{PASP }

\def\mnras{MNRAS }
\def\physrep{Physics Reports }
\def\iaucirc{IAUC }

\def\apjl{ApJL }

\title[Timing and Spectral Analysis of XTE~J0421+560/CI Cam]{Timing and Spectral Analysis of the Unusual X-Ray Transient XTE~J0421+560/CI Camelopardalis}
\author[E. S. Bartlett et al.]{E. S. Bartlett$^{1}$\thanks{E-mail:
e.s.bartlett@soton.ac.uk (ESB)}, J. S. Clark$^{2}$, M. J. Coe$^{1}$, M. R. Garcia$^{3}$ and P. Uttley$^{4}$ \\
$^{1}$School of Physics and Astronomy, University of Southampton, Highfield, Southampton, SO17 1BJ, United Kingdom\\
$^{2}$Department of Physical Science, The Open University, Walton Hall, Milton Keynes, MK7 6AA, United Kingdom\\
$^{3}$Harvard-Smithsonian Center for Astrophysics, 60 Garden Street, Cambridge, 02138, USA\\
$^{4}$Astronomical Institute ``Anton Pannekoek", University of Amsterdam, Postbus 94249, 1090 GE Amsterdam, The Netherlands}
\begin{document}

\date{Accepted 20XX Date. Received 20XX Date; in original form 20XX Date}

\pagerange{\pageref{firstpage}--\pageref{lastpage}} \pubyear{2012}

\maketitle

\label{firstpage}

\begin{abstract}
We present a detailed X-ray study of the 2003 \emph{XMM-Newton} observation of the High Mass X-ray Binary XTE~J0421+560/CI Cam. The continuum of the X-ray spectrum is well described by a flat power law ($\Gamma=1.0\pm0.2$) with a large intrinsic absorbing column ($N_H=(4.4\pm0.5)\times10^{23}$~cm$^{-2}$). We have decomposed the broad iron line into 3 separate components: Fe\textsc{i}-K$\alpha$, Fe\textsc{i}-K$\beta$ and Fe\textsc{xxiv-xxv}K$\alpha$. It is unclear how both neutral and almost fully ionised iron can exist simultaneously, however we suggest this could be evidence that the compact object is embedded in the circumstellar material. This doesn't appear to be consistent with the X-ray flux and spectrum of the source, which has remained essentially unchanged since the initial outburst. The iron abundance implied by the ratio of the neutral Fe-K$\alpha$ and Fe-K$\beta$ is compatible with solar. We search for lags in the neutral Fe-K$\alpha$ with respect to the continuum and find marginal evidence for a lag at $\sim$10~ks. We interpret this as the light crossing time of the torus which would suggest that the neutral iron is located at a radius of 10~AU. This result depends on several assumptions including the distance to the system, the inclination, the mass of system and the orbital period none of which are known with any great certainty. Better constraints on these system parameters and further observations of this system are required to confirm this result. We discuss the nature of this system in light of our results and place it in context with other binary B[e] stars.
\end{abstract}


\section{Introduction}

The X-ray transient XTE~J0421+560 was discovered by the All-Sky Monitor on-board RXTE during a strong outburst in 1998 by \citeauthor{Smith98} It brightened to a peak flux of $\sim$2 Crab within a few hours before rapidly decaying, reaching quiescence in under two weeks. Observations at different wavelengths immediately after the outburst revealed similar behaviour in the optical and radio bands with a peak flux density of almost 1~Jy at 8.30~GHz (e.g. \citealt{Clark00}). Initial radio observations suggested that there might be a superluminal expansion \citep{Hjellming98}, but subsequent analysis showed that the radio emitting region was expanding with a velocity of only $\sim$12,000~km~s$^{-1}$ \citep{mio.rup}. The X-ray spectra seen by \emph{SAX} \citep{Frontera98}, \emph{RXTE} \citep{Belloni99} and, more recently, \emph{XMM-Newton} \citep{Boirin02} are dominated by an emission feature at 6-7 keV attributed to the Fe-K$\alpha$ fluorescence line.

The optical counterpart of XTE~J0421+560 is CI Camelopardalis (CI Cam), a B0-2 supergiant B[e] star as classified by \citet{Clark99}, \citet{Hynes02} and \citet{Robinson02}, establishing this system as a High Mass X-ray Binary (HMXB). Common properties of stars exhibiting the B[e] phenomenon are the presence of forbidden emission lines in their optical spectra and a strong infra-red (IR) excess. This IR excess is attributed to a hot circumstellar dust shell \citep{Zickgraf86}. Until the launch of \emph{INTEGRAL}  CI~Cam was the only known sgB[e] companion star in a HMXB. \emph{INTEGRAL} has since identified another sgB[e]/XRB candidate, IGR~J16318-4848 \citep{Courv03}.

The circumstellar gaseous environment of CI Cam has been under much scrutiny since the 1998 outburst. \citet{Robinson02} propose a two-component wind from the sgB[e] star to explain their observations; a cool, low velocity ``iron'' wind and a hot, high velocity wind. The low velocity wind is dense, roughly spherical and extends to a radius of 13 to 50~AU. They note it is unclear as to how these winds with very different densities and velocities co-exist. Following a series of spectroscopic observations,\citet{Hynes02} also conclude that the emission lines must originate from several physically separate regions. Unlike \citeauthor{Robinson02}, they associate the H Balmer lines, He~\textsc{i} and Fe~\textsc{ii} lines with the same region; a disk, or equatorial outflow region, viewed nearly pole on. This is consistent with the two component stellar wind model for supergiant B[e] stars proposed by \citet{Zickgraf85,Zickgraf86}. 

Analysis of high resolution optical spectra by \citet{Miro02}, observed almost 3 years after the outburst, revealed clearly resolved emission line profiles, displaying a triple peak structure. They suggest that this implies an intermediate inclination angle of the circumstellar disk with respect to the line of sight. Recent interferometric work in the near infrared over an eight year period from the outburst by \citet{Thureau09} resolves the emitting region. They model this as a skewed elliptical Gaussian ring with semi-major axis $a=19.0\pm1.2$~AU (assuming a distance of 5~kpc) with an inclination angle of $i\simeq67^{\circ}$. \citet{Barsukova06} report a 19.41$\pm$0.02 day orbit period in the optical light curve of CI Cam, along with Doppler shifts in the He\textsc{ii}~4686~\AA{} line on the same time scale. The proposed period would put the semi-major axis of the binary orbit between 100-130~$R_\odot$: close to the surface of the star (based on mass and radii measurements of sgB[e] stars by \citealt{Zickgraf86}).

Unfortunately, distance estimates to XTE~J0421+560 are uncertain with a wide range of values being adopted in the literature. This is primarily due to the lack of reliable spectral diagnostics available to determine the the luminosity class of the mass donor. Consequently, estimates range from 1-17~kpc. This in turn means that the X-ray luminosity of the outburst is not well constrained and as a result the nature of the compact object of this binary system is still unclear. We critically discuss the distance and nature of the system in Section 5.

This paper reports on a 60~ks \emph{XMM-Newton} observation of XTE~J0421+560 during satellite revolution \#0588 on 2003 February 24. We discuss the data collected with the EPIC MOS \citep{Turner01} and pn \citep{Struder01} detectors and attempt to constrain the system geometry via spectral and temporal analysis. We also discuss the nature of the compact object and the system in light of these results and place it in context with other binary B[e] stars.

\section{Observations and Data Reduction}
\begin{table}
\centering
\caption{\emph{XMM-Newton} EPIC observations of XTE~J0421+560 on 2003 February 24}\label{table:obs}
\begin{tabular}{cccccc}
\hline\hline
Camera & Filter & Read out & \multicolumn{2}{c}{Observation} & Exp. \\
& & Mode & Start & End(UT) & (ks) \\
\hline
MOS1/2 & Medium & FF, 2.6~s & 12:34 & 05:50 & 59.5 \\
pn & Thin1 & FF, 73~ms & 12:56 & 05:50 & 60.3 \\
\hline
\end{tabular}
\end{table}Table \ref{table:obs} summarises the details of the EPIC observations. All detectors were operating in Full Window modes, giving a field of view of $\sim$30\arcmin diameter. The data were processed using the \emph{XMM-Newton} Science Analysis System v9.0 (SAS) along with software packages from \textsc{ftools} v6.8.

The MOS and pn observational data files were processed with \textsf{emproc} and \textsf{epproc} respectively. ``Single and double'' (PATTERN$\leq$4) pixel event patterns were selected for the pn detector. For the MOS instruments, ``single'' to ``quadruple ''(PATTERN$\leq$12) pixel events were selected. Source counts were extracted from a 30\arcsec radius region centered on XTE~J0421+560 and were compared with those from statistically identical background regions with extraction radii 60\arcsec. Whilst several periods of high background activity are present in the data, the source is brighter than the background for all but the final $\sim$13~ks of the observation. Hence we have not included this final section in the light curves reported here. The counts from the closest of the statistically identical background regions (located on either a neighbouring CCD (pn) or the same CCD (MOS1 and 2) as XTE~J0421+560) were used as the final background regions for the light curves. The background subtraction was performed using the \textsf{epiclccorr} task which also corrects for bad pixels, vignetting and quantum efficiency.

For the spectra, periods of high background activity were screened by removing any times when the count rate above 10~keV was \textgreater0.8 cts s$^{-1}$ for the MOS detectors and \textgreater1.5 cts s$^{-1}$ for the pn detector. Spectra were extracted using a 24\arcsec extraction radius as recommended by the SAS task \textsf{eregionanalyse}, which maximises the signal to noise ratio based on the source counts. ``Single and double'' pixel events (PATTERN$\leq$4) were accepted for the pn detector, all bad pixels and columns were disregarded (FLAG=0). For the MOS, ``single'' to ``quadruple ''(PATTERN$\leq$12) pixel events were selected with quality flag \#XMMEA\_EM.

Background spectra were taken from regions with a 60\arcsec radius away from the source on the same CCD. Several background regions were extracted and examined to ensure they were statistically identical. The region from which the background spectrum for the pn detector was selected has a similar distance to the readout node as the source region, as recommended by the ``EPIC status of calibration and data analysis'' document\footnote{http://xmm2.esac.esa.int/docs/documents/CAL-TN-0018.pdf}. For the MOS detectors, the same source region was used for both the MOS1 and MOS2 detectors. The area of source and background regions were calculated using the \textsf{backscal} task. Response matrix files were created for each source using the task \textsf{rmfgen} and \textsf{arfgen}. 

\section{Spectral Analysis}

The spectral analysis discussed in this work was performed using \textsf{XSPEC} \citep{Arnaud96} version 12.7.0. The spectra from the three EPIC detectors were fit simultaneously with the model parameters constrained to be identical across the three instruments. All models have an additional constant factor to account for the instrumental differences. All elemental abundances are set to the values of \citet{Wilms00} unless otherwise stated.\begin{figure} 
 \begin{center}
 \hspace{-18.5pt}
  \includegraphics[angle=90,width=0.51\textwidth]{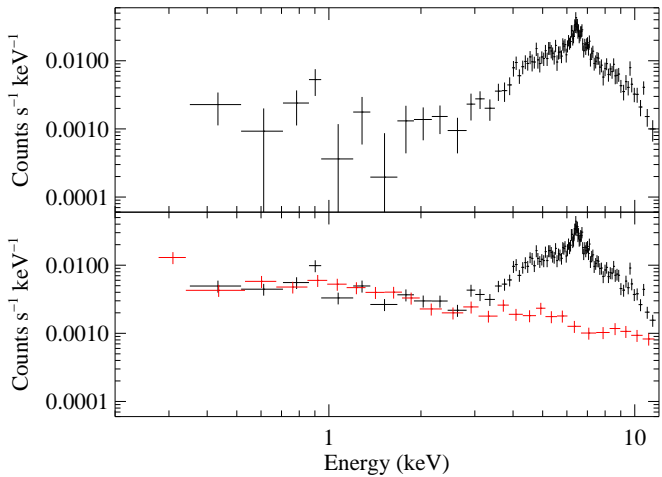}
 \end{center}
\caption{The 0.2--12.0~keV spectrum of XTE~J0421+560. Top panel shows the background subtracted spectrum, bottom panel shows the source spectrum before background subtraction (in black) along with the spectrum of the background (red). All spectra are normalised.}\label{fig:no_model}\end{figure}

Figure \ref{fig:no_model} shows the spectrum of XTE~J0421+560 both before and after background subtraction along with the background spectrum used. The spectrum of XTE~J0421+560 is characterised by heavy absorption and a large emission feature at $\sim$6.5~keV. \citet{Boirin02} report a soft (\textless2~keV) excess in their 2001 observation. Whilst there does appear to be evidence for a soft excess in our background subtracted spectrum, when the uncorrected source spectrum is compared with the background spectrum, the shape of the spectrum and value of the data points below 3~keV are almost identical. A visual inspection of the X-ray image shows no detection of the source below 3~keV. Whilst we cannot rule out the existence of a soft excess in the spectrum of XTE~J0421+560, we believe that this could be an artifact of the background subtraction and so do not include the \textless3~keV data points in our model fits.

The continuum of the spectrum was first fit by excluding the data around the iron line region. The absorption was modeled with two elements, a Galactic foreground component, $N_{H,Gal}$ fixed to $4.5\times10^{21}$~cm$^{-2}$ \citep{DL91} and a separate column density, $N_{H,i}$, intrinsic to the source which was allowed to vary. A simple absorbed power-law model (\emph{phabs*phabs*powerlaw} in \textsf{XSPEC}) yielded a respectable fit with a $\chi^2$ value of 135.6 for 128 degrees of freedom (dof; $\chi_r^2=1.06$). A partially covered power law model was also fit to the data (\emph{phabs*pcfabs*powerlaw} in \textsf{XSPEC}). In this model the underlying powerlaw is fractionally covered by an absorbing material with the uncovered emission absorbed by the interstellar medium. This did not improve the continuum fit with a $\chi^2$ value of 144.4 for 127 dof ($\chi_r^2=1.14$). We also fit the continuum with a power law partially covered by an ionised absorber (\emph{phabs*pcfabs*zxipcf} in \textsf{XSPEC}). This produces a statistically better fit than the partial covering model alone with a $\chi^2$ of 138.4 for 133 dof ($\chi_r^2=1.04$), but with unfeasible model parameters ($\Gamma=3.4_{-1.6}^{+1.3}$). Again, the covering fraction and ionisation parameter both suggest that an ionised partial covering model is unnecessary.

\begin{figure}
\vspace{10pt} 
 \begin{center}
  \includegraphics[width=0.50\textwidth]{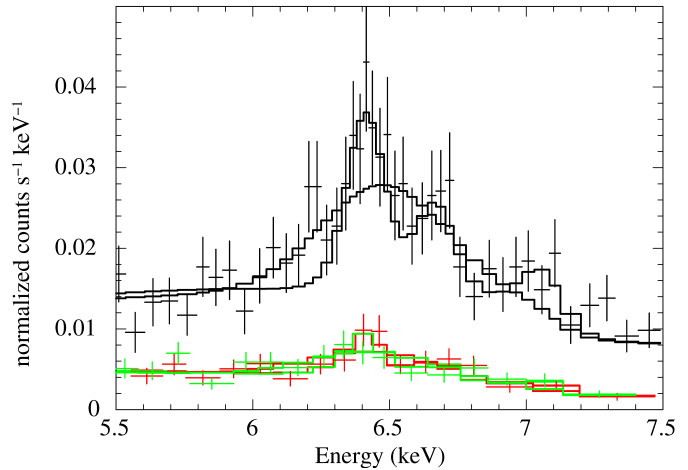}
 \end{center}
\caption{The EPIC spectrum of XTE~J0421+560 around the iron line along with the single and triple Gaussian models fit to the data.}\label{fig:Fe}\end{figure}

The region around the iron line was then included in the model fit. A single (unabsorbed), intrinsically narrow, Gaussian was added to the model with the energy of the line allowed to vary. The resulting fit has a K$\alpha$ line at $6.43\pm0.02$~keV. The fit has a $\chi^2$ of 196.7 ($\chi_r^2$ of 1.17 for 168 dof). A better fit is produced by allowing the Gaussian width ($\chi$ of 167.3; $\chi_r^2$ of 1.00 for 167 dof), with a broad iron line at $6.48\pm0.04$ with $\sigma=0.23_{-0.06}^{+0.08}$. Figure \ref{fig:Fe} shows the iron line region along with the best fit to the model and the broad iron line. Whilst statistically better than a single, narrow Gaussian, this fit has significant residuals and appears to be fitting the iron edge at 7.1~keV within the Gaussian. We do not believe this fit is physical.

A second intrinsically narrow Gaussian was added to the model at 6.67~keV, consistent with ionised Fe-K$\alpha$. This further reduced the $\chi^2$ to 171.0 ($\chi_r^2$=1.03 for 166 dof). Finally, a K$\beta$ line was added to the model fixed at 7.06~keV. The normalisation was initially constrained to be 0.13 of that of the K$\alpha$ line \citep{Kaastra93}. This reduced the $\chi^2$ to 164.7 ($\chi_r^2$=0.98 for 168 dof). When the normalisation and energy are unconstrained the line energy remains essentially unchanged ($E_{K\beta}=7.05_{-0.07}^{+0.08}$) and the ratio between K$\beta$ and K$\alpha$ rises to 0.3$\pm$0.2, consistent with the theoretical value.\begin{table*}
{\small
\centering
\caption{Best fit parameters for the \emph{phabs*((phabs*po)+Gaus+Gaus+Gaus)} model fits to the full 3.0--12.0~keV spectrum.  Errors, where reported, are the 90\% confidence level.}\label{table:line}
\begin{tabular}{ccccccccc}
\hline\hline\noalign{\smallskip}
 & & \multicolumn{2}{c}{Fe\textsc{i} K$\alpha$}  & \multicolumn{2}{c}{Fe\textsc{xxiv-xxv} K$\alpha$} & \multicolumn{2}{c}{Fe\textsc{i} K$\beta$}  & \\\noalign{\smallskip}
$N_{H,i}$ &$\Gamma$ & Energy & Flux & Energy & Flux & Energy & Flux & \multirow{2}{*}{$\chi^2$/dof}\\\noalign{\smallskip}
($\text{cm}^{-2}$) &  & (keV) & ($\text{ph}\:\text{s}^{-1}\:\text{cm}^{-2}$) & (keV) & ($\text{ph}\:\text{s}^{-1}\:\text{cm}^{-2}$) & (keV) & ($\text{ph}\:\text{s}^{-1}\:\text{cm}^{-2}$) & \\\noalign{\smallskip}
\hline\noalign{\smallskip}
$(5.6\pm0.6)\times10^{23}$ & $1.4\pm0.2$ & - & - & - & - & - & - & 280.1/170 \\\noalign{\smallskip}
$4.8_{-0.5}^{+0.6}\times10^{23}$ & $1.2\pm0.2$ & $6.43_{-0.03}^{+0.02}$ & $(6\pm1)\times10^{-6}$ & - & - & - & - & 196.7/168 \\\noalign{\smallskip}
$(4.5\pm0.5)\times10^{23}$& $1.1\pm0.2$ & $6.42_{-0.03}^{+0.01}$ & $(6\pm1)\times10^{-6}$ & $6.67_{-0.04}^{+0.03}$ & $(3\pm1)\times10^{-6}$ & - & - & 171.0/166 \\\noalign{\smallskip}
$(4.4\pm0.5)\times10^{23}$ & $1.0\pm0.2$ & $6.41_{-0.02}^{+0.03}$ & $(6\pm1)\times10^{-6}$ & $6.67_{-0.03}^{+0.02}$ & $(3.3_{-1.0}^{+0.9})\times10^{-6}$ & $7.05_{-0.07}^{+0.08}$ & $(2.0\pm0.9)\times10^{-6}$ & 161.0/164 \\\noalign{\smallskip}
\hline
\end{tabular}}
\end{table*}

Determining the significance of the iron lines in the spectrum is not straightforward. \citet{Protassov02} in particular warn against using the F-test to determine the significance of an emission line. Allowing the line energy to be free (as in our fitting) means that the models are no-longer nested and so the energy of the line cannot be incorporated as a free-parameter in the F-test. Instead, we need to allow for the fact that we have conducted a number of trials, $N$, over the energy range of the line. Conservatively, we estimate that the chance detection probability is at most $\sim1-(1-P)^N$ (see \citealt{Porquet04} for a full discussion). Since the line energies are within the instrument resolution of the expected values of 6.4~keV and 6.7~keV and not simply at some arbitrary redshift, the number of independent trials that we have effectively searched over is small. Here we adopt a value of 3 for $N$.

The data were fit with the simple absorbed power law model and the narrow iron lines added incrementally, as described above. They were then frozen at the best fit energy from the \emph{phabs*((phabs*po)+Gaus+Gaus+Gaus)} model and an F-test was performed after the addition of each line. As such each model is compared with the previous, simpler model. We also estimate the significance using the equivalent widths of the lines and their errors. The results of both of these methods are listed in Table \ref{table:signif}. Both the neutral and ionised K$\alpha$ lines appear to be clearly significant. There appears to be only a marginal detection of the expected K$\beta$ line, nonetheless the flux of the line is consistent with Solar metallicity.

\begin{table}
\centering
\caption{Estimates of the significance of the iron lines based on the two methods described in the text. Errors, where reported, are the 90\% confidence level.}\label{table:signif}
\begin{tabular}{lccc}
 \hline\hline\noalign{\smallskip}
F-Test & Fe\textsc{i} K$\alpha$ & Fe\textsc{xxiv-xxv} K$\alpha$ & Fe\textsc{i} K$\beta$\\
\hline\noalign{\smallskip}
F statistic & 79.3 & 30.4 & 10.1\\
Chance detection & \multirow{2}{*}{$2.2\times10^{-15}$} & \multirow{2}{*}{$3.8\times10^{-7}$} & \multirow{2}{*}{$5.2\times10^{-3}$} \\
probability & & & \\
 \hline\hline\noalign{\smallskip}
Equivalent width & Fe\textsc{i} K$\alpha$ & Fe\textsc{xxiv-xxv} K$\alpha$ & Fe\textsc{i} K$\beta$\\
\hline\noalign{\smallskip}
Eqwidth (eV) & $220_{-50}^{+90}$ & $100_{-40}^{+30}$ & $90\pm50$ \\
Significance & 7.24$\sigma$ & 4.11$\sigma$ & 2.96$\sigma$\\
\hline\noalign{\smallskip}
\end{tabular}
\end{table}

The continuum and line parameters for the best fit model are listed in Table \ref{table:line}.\begin{figure}
\centering
\hspace{-50pt}
 \includegraphics[angle=270,width=0.56\textwidth,]{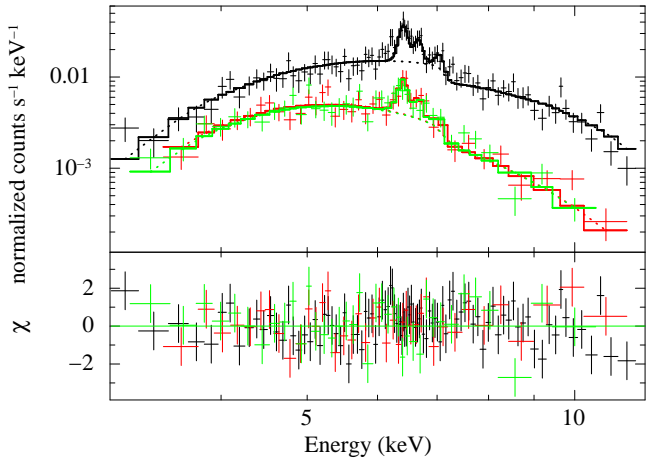}
\caption{EPIC-pn (black), EPIC-MOS1 (red) and EPIC-MOS2 (green) spectra, along with best ﬁt \emph{phabs*((phabs*po)+Gaus+Gaus+Gaus)} model to the data (solid line). Broken line shows the absorbed power law component.}\label{fig:bestfit}
\end{figure} Figure \ref{fig:bestfit} shows the best fit model to the spectrum. The final model fit has a photon index of $\Gamma=1.0\pm$0.2 and an intrinsic absorption of $N_H=4.4\pm0.5\times10^{23}$~cm$^{-2}$, with a $\chi^2$ of 161.6 ($\chi_r^2$=0.98 for 164 dof).

We also attempted to fit the thermal model of \citet{Ishida04}, derived from an \emph{ASCA} observation, to the data. In this model emission from two regions of hot diffuse gas are photoelectrically absorbed independently (\emph{phabs*mekal+phabs*mekal} in \textsf{XSPEC}) with the elemental abundances of \citet{angr}. The parameters were frozen to the best fit values of \citet{Ishida04}, with only the normalisation allowed to vary, and fit to the 0.5--10.0~keV spectrum. This allowed for a direct comparison. The resulting fit was poor with a $\chi^2$ of 1023.6 ($\chi^2_r$=5.62 for 182 dof), rising to 1040.6 ( $\chi^2_r$=6.09 for 171 dof) when only the 3.0--12.0~keV energy range was considered. Allowing the absorption and plasma temperature to vary improved the fit ($\chi^2$=308.2; $\chi^2_r$=1.7 for 178 dof) but with unfeasible model parameters (e.g.  $N_H\sim10^{26}$~cm$^{-2}$, kT$\sim$80~keV).

\section{Timing Analysis}

\begin{figure}
 \begin{center}
  \includegraphics[angle=90,width=0.5\textwidth]{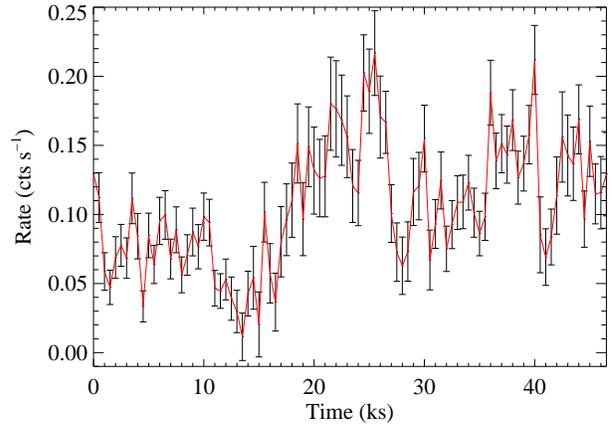}
 \end{center}
\caption{Background subtracted 3.0-12.0~keV light curve of XTE~J0421+560 from the EPIC-pn detector. The broken line shows the median count rate of the light curve.}\label{fig:entire}
\end{figure}

Figure \ref{fig:entire} shows the background subtracted 3.0--12.0~keV light curve for XTE~J0421+560, with a bin time of 500~s, from the EPIC-pn detector. The total net counts for the light curve, after background subtraction and excluding triple and quadruple pixel events, is $\sim$5000 counts. This corresponds to an average net count rate of 0.107$\pm$0.002~counts~s$^{-1}$, substantially brighter than the 2001 observation which found 0.024$\pm$0.002~counts s$^{-1}$ \citep{Boirin02}. The light curve has a fractional root-mean-squared variability amplitude of 0.37 (calculated using equation (10) of \citealt{Vaughan03}).

\begin{figure}
 \begin{center}
 \includegraphics[angle=90,width=0.5\textwidth]{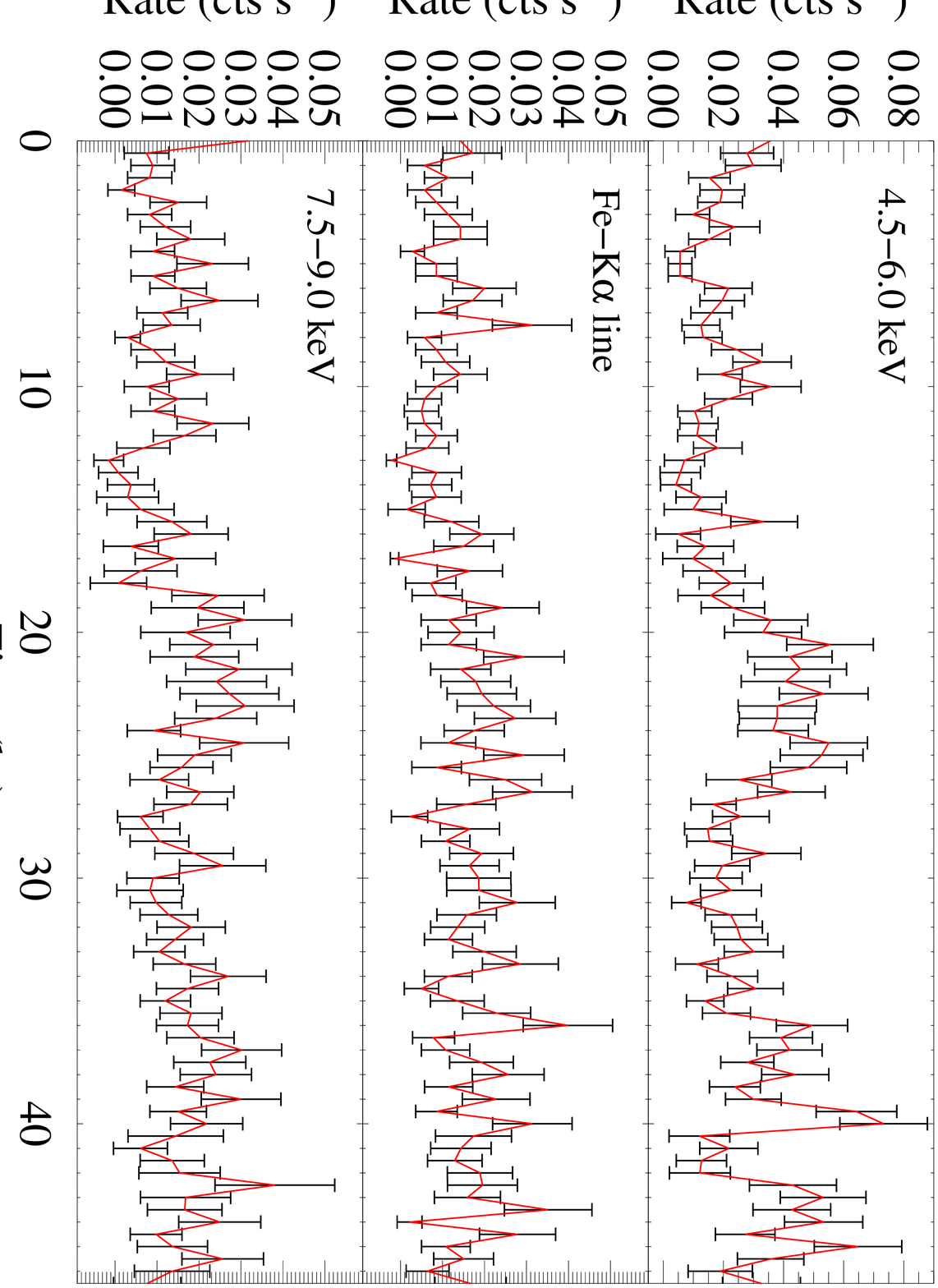}
\end{center}
\caption{light curves of the background subtracted low energy continuum (4.5--6.0~keV), the Fe-K$\alpha$ line (6.2--6.55~keV) and the high energy continuum (7.5--9.0~keV).}\label{fig:lcs}
\end{figure}Figure \ref{fig:lcs} shows the source light curve split into three different energy bands. The top panel shows the light curve for the lower energy continuum (4.5--6.0~keV), the middle panel shows the light curve in the range we have defined as the Fe-K$\alpha$ line from our spectral fits (6.20--6.55~keV) and the bottom panel shows the light curve for the higher energy continuum (7.5--9.0~keV). The Fe-K$\alpha$ line spans a much narrower range of energies than the continuum light curves. This is to avoid contamination from the hypothesised ionised K$\alpha$ line and the K$\beta$ line, as well as to minimise any contribution of the underlying continuum to the cross correlation function (CCF). Above $\sim$9.0~keV the signal to noise of the system reduces and so these data are not included in our analysis.

\begin{figure*}
 \begin{center}
 \includegraphics[width=0.85\textwidth]{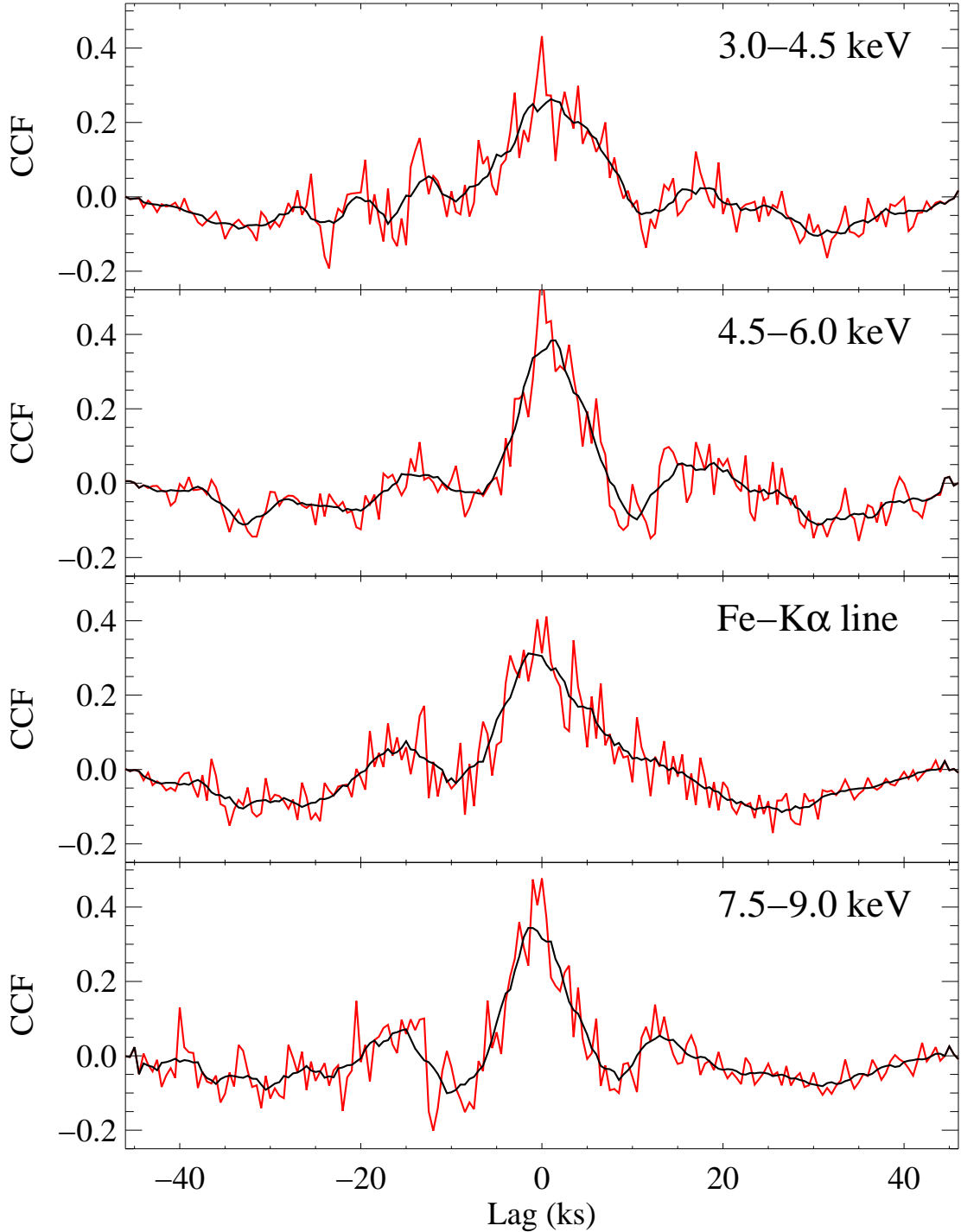}
\end{center}
\caption{The cross-correlation functions of the light curves. The continua light curves (3.0-4.5~keV, 4.5--6.0~keV and 7.5--9.0~keV) were cross-correlated with the sum of the remaining continua light curves whilst the Fe-K$\alpha$ (6.2-6.55~keV) light curve was cross-correlated with the sum of all the continua light curves. The solid black line is smoothed CCF.}\label{fig:CCF}
\end{figure*}

Cross-correlation analysis was performed on the three light curves mentioned above and an additional low energy continuum light curve ranging from 3.0--4.5~keV. The IDL routine \textsf{c\_correlate.pro} from the IDL Astronomy User's Library\footnote{http://idlastro.gsfc.nasa.gov/} was used throughout. The continuum light curves were correlated with the sum of the other 3 continuum light curves (i.e. the rest of the spectrum without the iron line complex). The Fe-K$\alpha$ line light curve was correlated with the sum of all the continuum light curves. Figure \ref{fig:CCF} shows the CCFs of the light curves with respect to the continua. The solid black lines are the same CCFs, smoothed using a moving average method in an attempt to minimise the noise. A positive time delay corresponds to a lag in the variation of the light curve with reference to the continua. The Fe-K$\alpha$ CCF appears to be more asymmetric than its continuum counterparts. More explicitly, the minimum present in the continua CCFs around $\sim$10~ks does not seem to be present.

Whilst the energy range of the Fe-K$\alpha$ line has been deliberately defined to keep continuum contamination at a minimum, inevitably there will be some contamination as there is no way to separate the line emission from the continuum emission at these energies. As such we expect to see some correlation of the light curves at zero lag. However, an asymmetry in the CCF is often interpreted as evidence for a lag (e.g. \citealt{Peterson98}) and so we explore the possibility that the shape of the CCF could be due a transfer function which includes some power at positive lags.

We define the ``asymmetry factor'', $A_{sym}$, of the CCFs to be the ratio of the difference of the half widths of the peak at tenth maximum, to the sum of the half widths. This gives a number between -1 and 1, with zero being perfectly symmetrical. A positive value of $A_{sym}$ indicates that the gradient of the peak is steeper on the negative lag side and vice versa for a negative value. The $A_{sym}$ values of the continua CCFs are all very similar, ranging from -0.04 to -0.08, suggesting they are essentially symmetrical. However, the $A_{sym}$ value for the Fe-K$\alpha$ CCF is 0.39, indicating a moderate level of asymmetry.

The method of \citet{Zhang02} was then used to see if the asymmetry seen in our Fe-K$\alpha$ CCF could be caused by Poisson noise. Monte-Carlo simulations were performed to assess the significance of this asymmetry. Ten thousand red-noise light curves were generated with a power law slope of -2.0 using the method of \citet{Timmer95} and the IDL routine \textsf{rndpwrlc.pro}. The light curves were initially simulated with a duration ten times longer than that of the actual data and were then cut down to the observed duration to account for red-noise leakage. The simulated light curve was then used to create four light curves with the statistical properties of the observed light curves, resulting in four identical light curves distinguishable only by their mean and standard deviation. Noise was then added to each data point of the simulated light curves by adding a Gaussian deviation with a mean of zero and a variance equal to that of the corresponding data point in the real light curve (similar to the method of \citealt{Uttley03}). Any negative data points were then set to zero. The three continuum light curves were then combined and cross-correlated with the fake Fe-K$\alpha$ line light curve in the exact same manner as the observed data. The asymmetry factor and peak value of the smoothed CCFs were then calculated.

Of the 10$^4$ CCFs generated in the Monte Carlo simulation, 303 were observed to have $A_{sym}>0.39$ suggesting a significance of $\sim$97\% or 2.1$\sigma$. This result remains at a similar significance (2.3$\sigma$) when a red noise power law index of -1.0 is considered. The asymmetry factors and peak values of the continuum CCFs all fall within the central 50\% of the distribution. Whilst this result is not formally significant, the low signal to noise of our data makes a convincing result unlikely.

\section{Discussion}

The continuum parameters of our best fit model are consistent with those reported by \citet{Boirin02}. The total 3.0-10.0~keV flux of our model is  $(1.36_{-0.05}^{+0.04})\times10^{-12}$~ergs~cm$^{-2}$~s$^{-1}$ corresponding to an X-ray luminosity of $L_{X(3-10keV)}=(4.1_{-0.2}^{+0.1})\times10^{33}$~ergs~s$^{-1}$ at a distance of 5~kpc. This is a slight increase on that seen in the previous \emph{XMM} observation ($L_{X(0.5-10keV)}=3.5\times10^{33}$~ergs~s$^{-1}$; \citealt{Boirin02}), however as there are no errors quoted on this value we cannot say whether this increase is significant. Whilst these luminosities do not cover the same energy range, as the system is undetected below 3~keV in our \emph{XMM} observation it is unlikely that the $L_{X(0.5-10keV)}$ will vary much from the luminosity reported above. The flux values used to calculate the luminosities are derived from the spectral fits and so are also model dependant. The light curve seems to contradict the results from the spectral fits to the data, suggesting a factor of 4.46$\pm$0.09 increase in the average count rate over the two year period between data sets.  A \emph{BeppoSAX} observation in 2000 put an upper limit on the X-ray luminosity of $L_{X(1-10keV)}$\textless$2.5\times10^{33}$~ergs~s$^{-1}$ \citep{Parmar00}. As such, the X-ray emission from the source does appear to have shown a moderate level of variability over the last few years. 

The improved signal to noise has allowed the apparently broad Gaussian, reported by \citet{Boirin02}, to be decomposed into three intrinsically narrow components, neutral Fe-K$\alpha$ and K$\beta$ and helium like Fe\textsc{xxv} K$\alpha$. It is unclear how both neutral and almost fully ionised iron can co-exist and only seems possible if the ionised line emission arises from a physically separate region from the neutral emission, possibly local to the compact object. This would require the compact object to be in the same plane as the equatorial outflows of the companion stars and seems inconsistent with the X-ray activity of the system. The iron abundance implied by the ratio of the neutral K$\alpha$ and K$\beta$ line fluxes is just consistent with solar within errors.

\begin{figure}
 \begin{center}
  \includegraphics[width=0.4\textwidth]{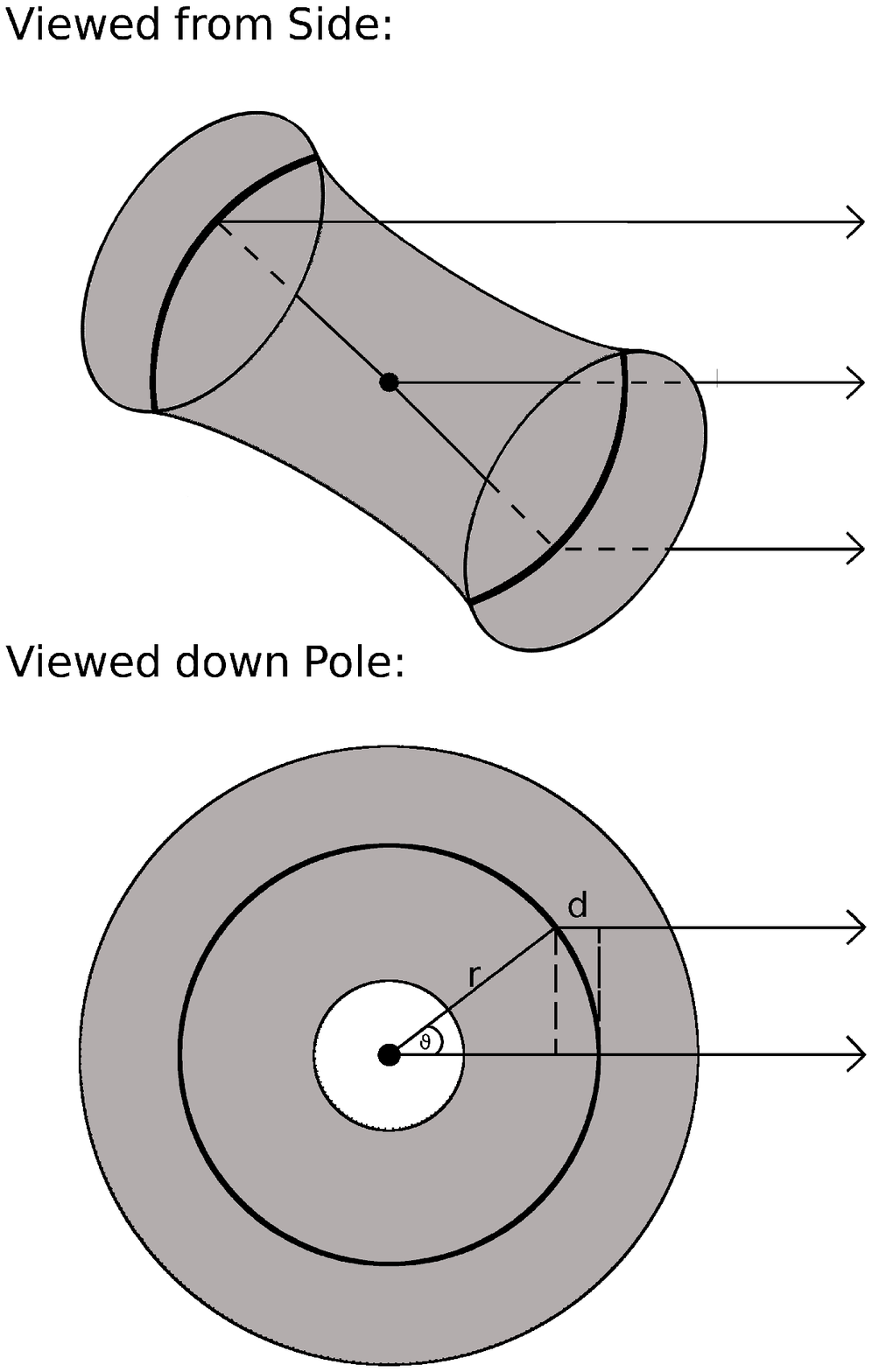}
 \end{center}
\caption{Sketch of the proposed geometry for CI Cam. The top figure shows the cross-section of the torus, whist the bottom figure shows the system as viewed down the pole. The thicker black ring at radius, \emph{r}, is the assumed location of the neutral iron. The central object in this case is the sgB[e] star. The position of the compact object is believed to be close to the surface of the optical star and thus lying well within the torus \citep{Barsukova06}.}\label{fig:torus}\end{figure}

Figure \ref{fig:torus} shows the current hypothesised geometry for CI~Cam proposed by \citet{Thureau09}, consisting of a binary system well within the torus. If we assume that the high column density originates from the leading (i.e. closest) edge of the surrounding torus (see Fig. \ref{fig:torus}) then the inclination, $i\sim67^\circ$, and semi-major axis, a=$19\pm1.2$~AU \citep{Thureau09}, give a lower limit to the height of the torus of 8.1$\pm$0.5~AU. This is highly dependent on both distance and inclination angle, neither of which are secure.

The K$\alpha$ fluorescence line of iron is already considered a powerful diagnostic tool for probing the central regions of both Galactic black hole candidate systems and active galactic nuclei \citep{Reynolds03}. A technique known as reverberation mapping involves searching for time dependent changes in the continuum, emitted by the primary X-ray source, and then looking for the echo, or ``reverberation'', in the emission lines located further out. The relationship between the continuum light curve and the emission line light curve is characterised by a geometry dependent ``transfer function''. Some idea of the spatial extent of the emission line region can be achieved by cross-correlating the light curves of the emission line, $L(t+\Delta\tau)$, and the continuum, $C(t)$. The function produced will be maximised at the lag, $\Delta\tau$, between the emission line and the X-ray continuum. If we assume that all the neutral iron is located in the torus at radius \emph{r}, that the compact object is in the centre of the torus (i.e. that the semi-major axis of the orbit is negligible compared to the semi-major axis of the torus) and that the X-ray emission is isotropic, then the range of lags seen is simply \begin{equation}
\Delta\tau=\frac{r}{c}(1-\cos\theta)
\label{eq:tau}
\end{equation}
where $\theta$ is the angle, between the line of sight and the X-rays incident on the torus (see Fig \ref{fig:torus}). A more in depth description along with the fundamental principles and assumptions of reverberation mapping is given by \citet{Bland82} and \citet{Peterson93}.

There appears to be evidence for an asymmetry in the CCF of iron K$\alpha$ energy range with the continuum, however much higher signal to noise data is needed to confirm this. If we consider the light curve of the Fe-K$\alpha$ energy range, $L(t+\Delta\tau)$, as the the sum of two lightcurves, $L_1(t+\Delta\tau)$ representing the line emission and $L_2(t)$ representing the continuum ``underneath" the iron line in the same energy range, then the CCF produced can be expressed as the sum of two CCFs:{\small\begin{align*}
F_{CCF}(\tau)&=\int_{-\infty}^\infty L(t+\Delta\tau)\,C(t)\,dt \\
&=\int_{-\infty}^\infty (L_1(t+\Delta\tau)+L_2(t))\,C(t)\,dt \\
&=\int_{-\infty}^\infty L_1(t+\Delta\tau)\,C(t)\,dt +\int_{-\infty}^\infty L_2(t)\,C(t)\,dt\tag{2}\label{eq:sum}\end{align*}} The first component of equation \ref{eq:sum} represents the \emph{un-normalised} CCF of the pure line emission (i.e. the CCF is not normalised by the root mean square of the light curve in each band) and the second component the \emph{un-normalised} CCF of the underlying continuum. If this asymmetry is real and the continuum behaviour is simple and well represented by the CCF of the 7.5-9.0~keV light curve with the continuum (chosen as it is these photons which will be directly promoting the K-shell electrons into the L-shell), then subtracting the un-normalised 7.5-9.0~keV CCF from the the un-normalised Fe-K$\alpha$ CCF will leave the CCF of the pure Fe-K$\alpha$ emission with the continuum.

Figure \begin{figure}
\begin{center}
 \includegraphics[angle=90,width=0.5\textwidth]{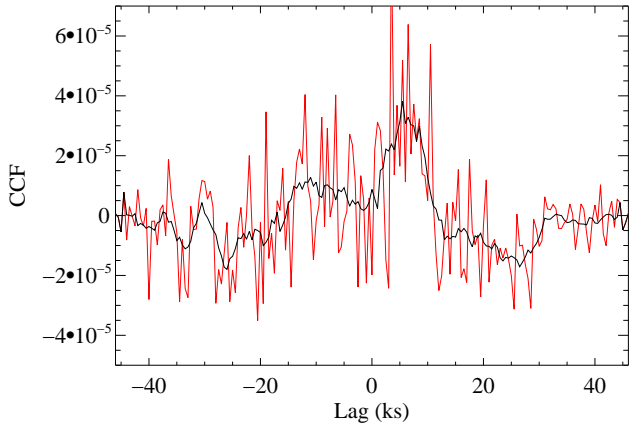}
\end{center}
\caption{The CCF produced when the un-normalised 7.5-9.0~keV CCF is subtracted from the un-normalised Fe-K$\alpha$ CCF.}\label{fig:subtraction}\end{figure}
\ref{fig:subtraction} shows the result of subtracting the un-normalised 7.5-9.0~keV CCF from the the un-normalised Fe-K$\alpha$ CCF (consistent with the mathematical proof above). The poor signal to noise is in part a result of the 0.34 fluorescence yield of iron \citep{Bamb72}; only $\sim$1 in 3 of the incident photons with energies \textgreater6.4~keV will result in a K$\alpha$ photon. However, there does appear to be a distinct peak at 5-10~ks.

The most likely values of $\tau$ from equation \ref{eq:tau} are $\sim0$ and $\sim2r$, where \emph{r} is the radius of the torus. If the peak is interpreted as the lag at $2r$ then this would indicate that the neutral iron is located at a radius of $\sim$10~AU from the centre of the torus. This result relies on several assumptions and parameters which are by no means secure. The derived semi-major axis of the orbit (from the period of \citealt{Barsukova06}) depends on the cubed root of total mass of the system. As such large changes in the total mass of the system only produce small alterations in the semi-major axes. If we consider a companion star of mass $50\pm30~M_\odot$ and a compact object with a mass $5\pm5~M_\odot$ (consistent with observed neutron star and stellar mass black hole mass estimates), then the semi-major axis is $0.5\pm0.1$~AU. This is a small fraction of the semi-major axis of the torus and so the approximation that the compact object is at the centre of torus seems valid. However, in reality the neutral iron is unlikely to be located in an infinitesimally thin ring in the torus, it seems more likely that there exists an inner radius at which neutral iron can form. In this case a lag would also have a more complex density dependence than the one that is considered here. Both these factors will serve to broaden the peak seen at $2r$, and so the derived radius is likely the radius where the neutral iron density is at a maximum.

\subsection{The nature of CI Cam}

A significant obstacle to understanding the nature and behaviour of CI Cam is the uncertainty in the distance, with estimates ranging from $\sim1.1$~kpc \citep{Barsukova06} to \textgreater10~kpc \citep{Robinson02}. The consequential uncertainty in the luminosity impacts on the nature of both the mass donor and the accretor, with the former authors favouring a B4\textsc{iii-v} + white dwarf model and the latter a sgB[e] star + relativistic accretor (neutron star or black hole). In support of the former interpretation, \citet{Ishida04} infer the presence of a white dwarf from analysis of the \emph{ASCA} X-ray spectrum. This was fit with an optically thin, thermal hard X-ray model, characteristic of cataclysmic variables which contain a white dwarf. Our analysis (Section 3) clearly favours the power law model, with similar parameters to those reported for the other known sgB[e] HMXB, IGR~J16318-4848 \citep{Matt03b,Barragan09}, which \citet{Filliatre04} claim hosts a neutron star based on the ratio of the X-ray and radio flux (though they note that the unique environment means they are cautious about applying such a relationship). The photon index reported is also consistent with those found in neutron star Be/X-ray binaries by \citeauthor{Haberl08} (0.6-1.4; 2008).

\subsection{Comparison to  other B[e] supergiants}

To date only two other supergiant B[e] stars have been detected as X-ray sources; the HMXB IGR~J16318-4848 (e.g. \citealt{Filliatre04, Chaty12}) and the cluster member Wd1-9 (\citealt{Skinner06, Clark08}). IGR J16318-4848 shows a striking similarity to CI~Cam, with a quiescent X-ray spectra best fitted with an absorbed, truncated power law and a strong Fe-K$\alpha$ line. There is no evidence for the ionised iron line in the higher resolution \emph{Suzaku} data of IGR~J1631-4848 \citep{Barragan09} suggesting that this could be unique to CI~Cam. The X-ray spectrum of Wd1-9 instead demonstrates a spectrum consistent with emission from an optically thin thermal plasma ($kT\sim 2.3$keV; \citealt{Clark08}). This is suggestive of emission from a colliding wind binary rather an accreting binary, consequently we do not discuss this source further.

Uncertainties in the distances to both CI Cam and IGR J16318-4848 mean that it is difficult to compare quiescent luminosities, but our results suggest that they appear broadly comparable (to within a factor of a few) under the assumption that CI Cam is located at $\sim$5~kpc and IGR J16318-4848 is within the 1.6-4~kpc range favoured by \citet{Chaty12} (\citealt{Matt03b}, Sect. 3) implying that the quiescent fluxes in both CI Cam and IGR J16318-4848 arise via a similar mechanism. Both sources have undergone  X-ray ``flares'' with current temporal sampling suggesting that these events are rare and aperiodic; only two flares separated by $\sim 9$~yrs have been associated with IGR J16318-4848 (1994 and 2003; \citealt{Courv03, Murakami03}) and just the one event in 1998 with CI Cam (noting that photometric observations between 1989-92 also show no evidence for flaring in this period; \citealt{Clark00}). Moreover the peak luminosity also differs between the systems; with \citeauthor{Filliatre04} suggesting $L_{X}\sim2.6\times10^{36}$ergs~s$^{-1}$ for IGR~J163818-4848 (assuming a distance of 4 kpc) compared to $L_{X}\sim3\times10^{38}$ergs~s$^{-1}$ for CI Cam (e.g. \citealt{Hynes02}).

Both CI Cam and IGR J16318-4848 appear to be rather luminous supergiants, with comparable bolometric luminosities. Both sources also support a mid-IR excess attributed to circumstellar gas. Although different modeling assumptions (optically thin/thick emission, spherical or disc geometry and composition) make direct comparison of the physical  properties (e.g. inner radius, dust mass) of the dusty components difficult. The dust temperature at the inner radius of the envelope appears rather high in both cases ($\sim 767$K for IGR J16318-4848 \citep{Chaty12} and $\sim 1550$K for CI Cam \citep{Thureau09}). The geometry  of the dusty envelope has been unambiguously determined for CI Cam by \citet{Thureau09}, who showed that it resides in a ring or torus with inner radius $\sim8$~AU (at 5~kpc). No conclusions as to the dust geometry in IGR J16318-4848 were drawn by \citet{Moon07}, although by analogy to other B[e] stars \citet{Chaty12} adopt a disc like geometry with inner radius $\sim1$~AU/kpc. If the $\sim$19.3~day orbital period \citet{Barsukova06} propose for CI Cam is confirmed, the  binary would also lie interior to the dusty disc mapped by \citet{Thureau09}. Unfortunately no orbital period is known for IGR J16318-4848.

\section{Conclusions}

We have presented a detailed analysis of the spectral and timing properties of XTE~J0421+560/CI~Cam. The continuum of the spectrum is well described by a heavily absorbed ($N_H=(4.4\pm0.5)\times10^{23}\text{cm}^{-2}$ power law with photon index consistent with those seen in neutron star Be/X-ray binaries ($\Gamma=1.0\pm0.2$). We also attempted to fit the white dwarf model of \citet{Ishida04} but could not achieve an acceptable fit or sensible model parameters. The apparently broad iron line has been decomposed into three, intrinsically narrow lines, neutral Fe-K$\alpha$ and K$\beta$ and almost completely ionised Fe\textsc{xxv} K$\alpha$, suggesting multiple emitting regions. The quiescent spectrum seems to point to a neutron star or black hole accretor, however this is inconsistent with the X-ray activity of the system and we cannot suggest a mechanism for the large X-ray flare seen in 1998.

Further observations of XTE~J0421+560/CI~Cam of increased duration are required to confirm or refute the presence of the lag suggested in this paper. Better quality data could be used to solve the transfer equation and could be used to explore the possibility of a lag between the neutral and ionised Fe-K$\alpha$ emission, both of which are beyond the scope of this work. Better constraints on other system parameters, such as distance, inclination and system mass, would also help to constrain the geometry of this unusual system using this method.

\bsp

\label{lastpage}

\end{document}